\documentclass[orivec,british,article,runningheads]{llncs}

\usepackage{mathtools}
\usepackage{subfig}
\usepackage{upgreek}
\usepackage{xspace}
\usepackage{mathpartir}
\usepackage{courier}
\usepackage[dvipsnames,usernames]{xcolor}
\definecolor{deepred}{rgb}{0.6,0.0,0.0}
\definecolor{deepblue}{rgb}{0.0,0.0,0.5}
\usepackage{cite}
\usepackage{listings}
\usepackage{proof-dashed}

\usepackage[utf8]{inputenc}
\usepackage{amsmath}
\usepackage{amssymb}
\usepackage{babel}
\usepackage[pdftex]{graphicx}
\usepackage{ifthen}
\usepackage{verbatim}
\usepackage{siunitx}
\usepackage{array}
\usepackage{xtab}
\usepackage{booktabs}
\usepackage{enumitem}
\usepackage{mod}

\usepackage{listings}
\usepackage{float}
\usepackage{longtable}
\usepackage[T1]{fontenc}
\usepackage{lmodern}
\usepackage{colortbl}
\usepackage{pgfplots}

\tikzset{hnodeNormal/.style={draw,ellipse,minimum height=12, minimum width=17,inner sep=1}}
\tikzset{tnode/.style={draw,circle,fill=black,inner sep=0,minimum size=3}}
\tikzset{hedge/.style={draw,rectangle,minimum height=12, minimum width=17,inner sep=1}}
\tikzset{edge/.style={->,>=stealth',thick}}
\tikzset{tedge/.style={->,>=stealth'}}

\newlength{\testwd}
\newcommand{\fitpic}[1]{
	\settowidth{\testwd}{\includegraphics{#1}}
	\message{#1 width=\the\testwd, page=\the\textwidth}
	\ifthenelse{\lengthtest{\testwd>\textwidth}}{
	\noindent\includegraphics[width=\textwidth]{#1}}{
	\includegraphics{#1}}
}

\newcommand{\summaryRuleSide}[1]{
	\begin{minipage}{0.28\textwidth}
		\begin{center}
			\fitpic{#1}
		\end{center}
	\end{minipage}
}

\newcommand{\summaryRuleSpan}[1]{
	\begin{center}
	\begin{tikzpicture}[thick,
		node distance=20pt,
		vertex/.style={draw},
		normal/.style={->,>=triangle 45} 
	]
	\node[vertex,label=above:$L$](L)				{\summaryRuleSide{#1_L}};
	\node[vertex,label=above:$K$](K)[right=of L]	{\summaryRuleSide{#1_K}};
	\node[vertex,label=above:$R$](R)[right=of K]	{\summaryRuleSide{#1_R}};
	\draw[normal](K) to (L);
	\draw[normal](K) to (R);
	\end{tikzpicture}\\
	\end{center}
}

\lstdefinelanguage{pseudo}{morekeywords={function,return,for,if,while,end,each,in,add,to,else,do,length}}
\allowdisplaybreaks

\newcommand{\piIn}[2]{#1(#2)}
\newcommand{\piOut}[2]{\overline{#1}[#2]}
\newcommand{\piPar}[0]{\mathop{|}}

\newcommand{\modCalc}[0]{\texttt{EpiM}\xspace}
\newcommand{\MOD}{\texttt{M{\O}D}\xspace}

\newcommand{\sysPi}{\mathcal{R}_{\pi}}
\newcommand{\sysGc}{\mathcal{R}_{gc}}
\newcommand{\sysM}{\mathcal{R}_{m}}
\newcommand{\sysR}{\mathcal{R}_{\pi}^{\phi}}

\newcommand{\str}[1]{\texttt{#1}}
\newcommand{\iso}{\ensuremath{\cong}}
\newcommand{\moreGeneral}{\ensuremath{\succeq}}
\newcommand{\moreSpecial}{\ensuremath{\preceq}}
\newcommand{\unifiable}{\ensuremath{\overset{u}{=}}}

\newcommand{\figname}[0]{Fig.\ }
\newcommand{\secname}[0]{Sec.\ }

\usepackage{marvosym}

\begin{document}
\title{
A Graph-Based Tool to Embed the $\pi$-Calculus into a Computational DPO Framework
}
\titlerunning{A Graph-Based Tool to Embed the $\pi$-Calculus.}
\author{
Jakob L.\ Andersen\inst{1} \and Marc Hellmuth\inst{2} 
\and Daniel Merkle\inst{1,3}\and~\\ 
Nikolai Nøjgaard\inst{1,2} \and
Marco Peressotti\inst{1}
}
\authorrunning{Andersen et al.} 
\institute{
Department of Mathematics and Computer Science, University of Southern
Denmark, Odense M, DK\\
\and
Institute of Mathematics and Computer Science, University of
Greifswald, DE
\and
Harvard Medical School, Department of Systems Biology, Boston, MA, US
}

\maketitle 

\begin{abstract}
  Graph transformation approaches have been successfully used to
  analyse and design chemical and biological systems. Here we build on
  top of a DPO framework, in which molecules are modelled as typed attributed
  graphs and chemical reactions are modelled as graph
  transformations. Edges and vertexes can be labelled with first-order
  terms, which can be used to encode, e.g., steric information of
  molecules.  While targeted to chemical settings, the computational
  framework is intended to be very generic and applicable to the
  exploration of arbitrary spaces derived via iterative application of
  rewrite rules, such as process calculi like Milner’s
  $\pi$-calculus. To illustrate the generality of the framework, we
  introduce \modCalc: a tool for computing execution spaces of
  $\pi$-calculus processes. \modCalc encodes $\pi$-calculus processes as
  typed attributed graphs and then exploits the existing DPO framework
  to compute their dynamics in the form of graphs where nodes are
  $\pi$-calculus processes and edges are reduction steps. \modCalc takes
  advantage of the graph-based representation and facilities offered
  by the framework, like efficient isomorphism checking to prune the
  space without resorting to explicit structural equivalences. \modCalc is
  available as an online Python-based tool.

\end{abstract}

\section{Introduction}

Graph transformation approaches have been shown to provide formalisms
that elegantly facilitate the construction of reaction rules in
organic chemistry and biology \cite{bionetgen,kappa,andersen2016software}. Many of
these frameworks, while aimed at modelling organic chemistry and biology, are
constructed as generic foundations which can be used to model in a wider variety of domains.

The framework focused on in this paper is \MOD and is traditionally used to
model organic chemistry \cite{andersen2016software,Andersen17}. Here, graphs represent
molecules, transformation rules specify how molecules can interact,
and direct derivations represent concrete chemical reactions.
Specifically, graphs transformations are in \MOD
performed with the double-pushout (DPO) approach, with injective graph morphisms \cite{ehrig1973graph,habel2001double}.
A higher-level ``strategy framework'' then allows for programming the sequence of rule to apply on sets of graphs,
while performing the necessary graph isomorphism checks \cite{strat:14}.
Additionally, graphs in \MOD are labelled with first-order terms which allows
for the specification of even more abstract rules with named variables as attributes.

In this paper we will use the foundational framework of \MOD to implement a
tool for computing execution spaces of $\pi$-calculus processes. $\pi$-calculus
is a process algebra that, like all process algebras, is concerned with the
problem of formally modelling concurrent systems
\cite{milner1999pibook,sangiorgi2001pibook}. A process in 
$\pi$-calculus represents a ``unit'' of computation. In a concurrent system
several of such processes can run concurrently and communication between
processes occurs via channels on which channel names are sent. Given a set of
processes, all running concurrently, the execution space is then all possible
state transitions up to structural congruence.

The modelling of $\pi$-calculus as a graph transformation system is not a new
concept \cite{gadducci2007graph,konig2000graph}. Similarly, tools already
exist that simulates $\pi$-calculus \cite{bog2006tool,Victor102920}. Few
tools, however, simulates the execution space of a process up to structural
congruence, and no tools, to our knowledge, does this by using established graph
transformation concepts such as the DPO approach.

Here, we present the tool \modCalc, which is a Python-based library that
embeds the language of $\pi$-calculus into the framework of \MOD. It allows the
modelling of processes as simple Python expressions which are then encoded into
graphs.  The
execution space of processes is modelled as a set of transformation
rules applied on their corresponding graph encodings. Structural congruence is
checked using efficient graph isomorphism checking of graphs labelled with
first-order terms provided by \MOD. The encoding and the simulation of process
communication via graph transformations are based on the results established 
in \cite{gadducci2007graph}, with the notable modification that our encoding
obtains simple labelled graphs, while their model is concerned with directed
typed hypergraphs.
A web front-end of \modCalc is provided at \url{http://cheminf.imada.sdu.dk/epim}.

\section{Preliminaries}

\subsection{The $\pi$-calculus}\label{sec:pi-calc}
The $\pi$-calculus is a mathematical model of concurrent interacting processes ($P$, $Q$, $R$, \dots) that communicate using names ($x$, $y$, $z$, \dots). A name represents a communication channel or session, and can be sent and received as a message \cite{milner1999pibook,sangiorgi2001pibook}.
Process terms are given by the following grammar:
\\[\abovedisplayskip]%
\begingroup\sbox0{\qquad$P, Q \Coloneqq {} $}%
\sbox1{$P + Q$}%
\xentrystretch{-0.1}%
\begin{xtabular}{%
		>{\raggedleft\(}m{\wd0}<{\)}%
		>{\raggedright\(\begingroup}m{\wd1}<{\endgroup\)}%
		>{\quad\it}l<{}}
	P, Q \Coloneqq {} & \piOut{x}{y}.P & output name $y$ on channel $x$ and continue as $P$\\
	\mid {} & \piIn{x}{y}.P & input a name on $x$, bind it to $y$ and continue as $P$\\
	\mid {} & P \piPar Q & run processes $P$ and $Q$ in parallel (parallel composition)\\
	\mid {} & P + Q & run either $P$ or $Q$ (choice)\\
	\mid {} & (\upnu x)P & bind $x$ in $P$ (restriction)\\
	\mid {} & 0 & terminated process
\end{xtabular}\endgroup%
\\[\belowdisplayshortskip]%
The sets $fn(P)$ and $bn(P)$ of free and bound names in a process $P$ are defined as expected as well as $\alpha$-conversion.

Intuitively, process terms that differ solely on the order of parallel composition $\piPar$, sums $+$, and restrictions $\upnu$ represent the same process. This intuition is formalised by \emph{structural equivalence}, i.e., the least relation $\equiv$ on process terms that is a congruence w.r.t.\ the grammar above and closed under the Abelian laws for $\diamond \in \{\piPar,+\}$ under $0$: 
\[
	P \diamond Q \equiv Q \diamond P
	\qquad
	(P \diamond Q) \diamond R \equiv P \diamond (Q \diamond R)
	\qquad
	P \diamond 0 \equiv 0
\]
and under the distributivity laws for restriction:
\[
	(\upnu x)(\upnu y)P \equiv (\upnu y)(\upnu x)P
	\quad
	(\upnu x)0 \equiv 0
	\quad
	(\upnu x)(P \piPar Q) \equiv (\upnu y) P \piPar Q \text{ for } x \notin fn(Q)
	\text{.}
\]

For the sake of simplicity, we adopt the common assumption that choices are always guarded (i.e.~every branch in a sum $+$ is either a choice or a communication): in the sequel we assume the grammar $P,Q \coloneqq M \mid P \piPar Q$ where $M,N \coloneqq 0 \mid \piIn{x}{y}.P \mid \piOut{x}{y}.P \mid M + N$ \cite{gadducci2007graph}.

\def\rname#1{\textsc{\small #1}}
The semantics of process terms is given by the reduction relation $\to$, i.e., the least relation closed under the rules below:
\begin{gather*}
\infer[\rname{Com}]
	{(\piIn{x}{y}.P + M) \piPar (\piOut{x}{z}.Q + N)
	 \to 
	 P\{y/z\} \piPar Q}
 	{}
\\
\infer[\rname{Str}]
	{P \to Q}
	{ P \equiv P' 
	& P' \to Q' 
	& Q' \equiv Q}
\qquad
\infer[\rname{Res}]
	{(\upnu x) P \to (\upnu x) Q}
	{P \to Q}
\qquad
\infer[\rname{Par}]
	{P \piPar Q \to P' \piPar Q}
	{P \to P'}
\end{gather*}
Rule \rname{Com} models synchronous communication between two processes, possibly under a non-deterministic context (subterm $M$ may simply be $0$). Rule \rname{Str} ensures structurally equivalent processes have the same behaviour. Rule \rname{Par} models the (interleaved) execution of parallel components. Rule \rname{Res} allows execution under a restriction.

\subsection{Graphs, Transformation Rules, and the \MOD framework}
In this section we will be giving a very brief overview of the \MOD framework
and refer to \cite{andersen2016software} for a full overview.
Encodings of processes will be represented as simple labelled graphs, where
vertices and edges of graphs are labelled with first-order terms.  
We use the common definitions of graph and monomorphisms for labelled graphs,
where we require a most general unifier to exist for the set of term mappings induced by the underlying graph morphisms.
For full details see Appendix \ref{app:graph-morphisms}.

Graph transformation rules are modelled in the traditional DPO framework with
injective morphisms. See \cite{habel2001double} of an overview of the DPO
approach. The span of a rule, $L \leftarrow K
\rightarrow R$ is illustrated like in \figname \ref{fig:reduction-rule}.  
Using first-order terms as labels is useful, since it allows us
to specify abstract graph transformation rules such as \figname
\ref{fig:reduction-rule}, where any term leading with an underscore, e.g.,
$\_X$, specifies variable term.

The application of graph transformation rules on some input graph, is done by
using a strategy framework, where a specific order in which the graph
transformation rules should be applied can be specified.
The direct derivations obtained from applying a given strategy on an input graph is illustrated by
a derivation graph. A derivation graph is a directed graph, where vertices
are graphs obtained from rule applications and where an edge $(G, H)$, represents
a direct derivation $G \Rightarrow H$. Any pair of graphs in the derivation
graph are non-isomorphic, and as a result the closure up to isomorphism of a
given strategy can be automatically computed by \MOD. See \figname
\ref{fig:reduction-dg-clean} for an example.

\section{Encoding $\pi$-calculus}
\begin{figure}[tbp]
	\subfloat[]{
			\includegraphics[width=0.5\textwidth]{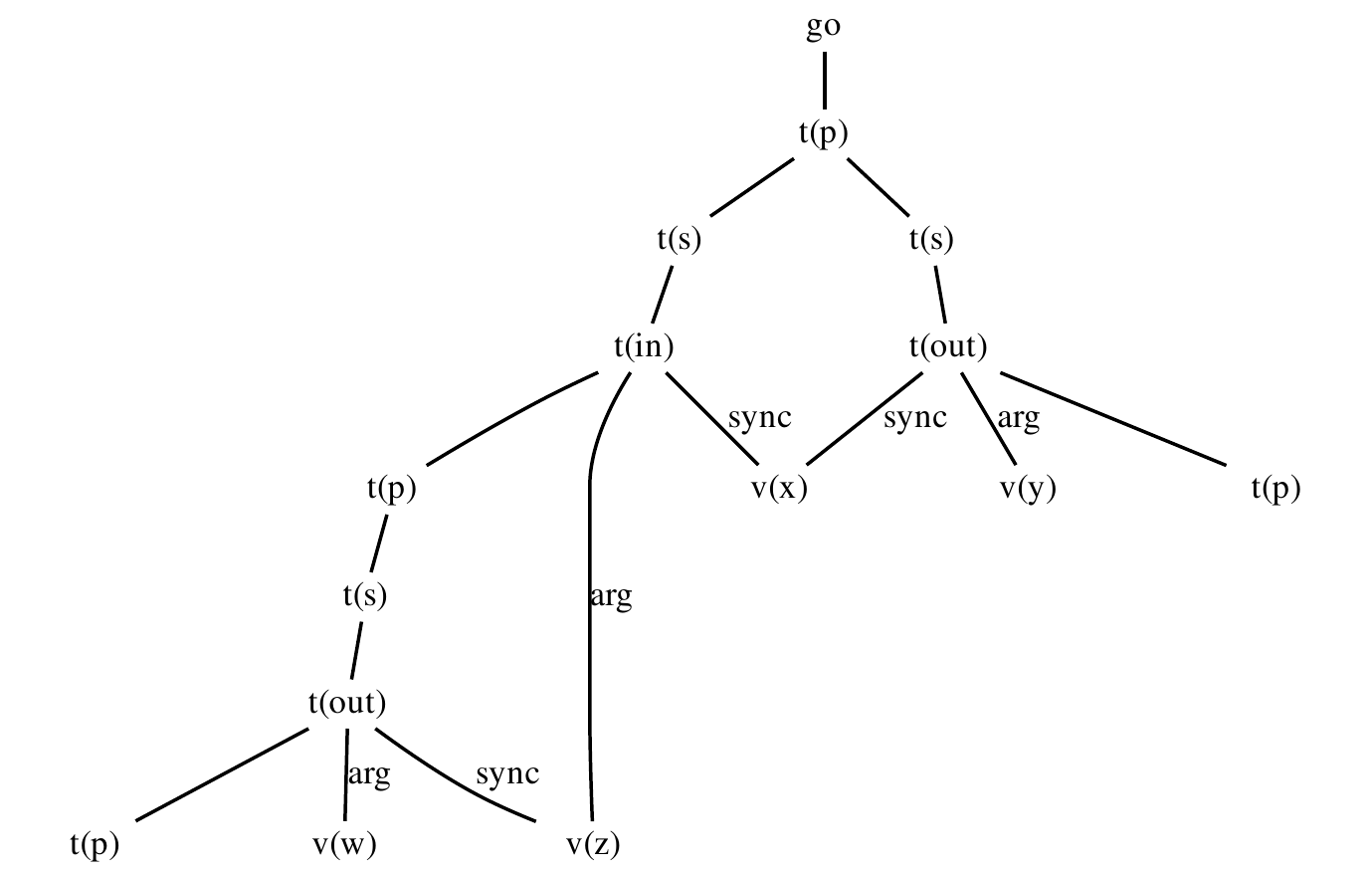}
	\label{fig:syntax-tree-nfp}
		}
	\subfloat[]{
		\includegraphics[width=0.5\textwidth]{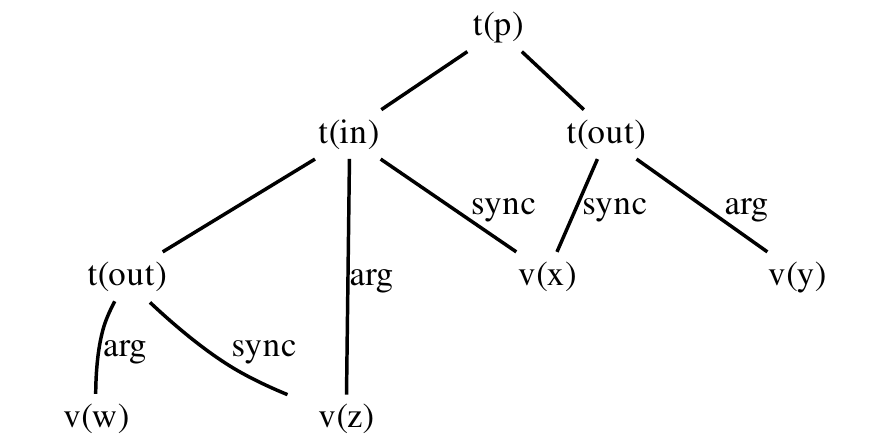}
		\label{fig:syntax-simplified}
		}
	\caption[]{
		\subref{fig:syntax-tree-nfp} The encoded process $[P]$ for $P = \piIn{x}{z}.\piOut{z}{w}\piPar \piOut{x}{y}$.
		\subref{fig:syntax-simplified} The simplified version of $[P]$.
	}
\end{figure}

Suppose we wanted to model the process $P = \piIn{x}{z}.\piOut{z}{w}\piPar
\piOut{x}{y}$ as a graph. To encode $P$ in \modCalc we can write
the following:
\begin{lstlisting}[language=Python]
x, y, z, w = names("x y z w")
p1 = Process().input(x, z).output(z, w)
p2 = Process().output(x, y)
P = p1 | p2
G = P.encode()
G.print()
\end{lstlisting}
Line 1 defines the names to be used in $P$. Line 2--3 defines the behaviour of
each subprocess in $P$, while line $4$ defines $P$ as the parallel composition
between them. Finally, line $5$ encodes $P$ as a graph that is printed in line
$6$. The resulting graph is illustrated in \figname \ref{fig:syntax-tree-nfp}.
The encoding is based on \cite{gadducci2007graph} modified to work on
simple labelled graphs instead of typed directed hypergraphs.

We denote the graph encoded by a process $P$ as $[P]$. The overall topology of
$[P]$ can be thought of as tree-like, describing the behaviour of $P$, just like
a syntax tree for a context-free grammar, but where names are shared between
operators. Since \MOD works on undirected graphs, $[P]$ is equipped with a
special vertex with the label $go$, representing the ``root'' of the tree. This
will be useful later, to be able to specify the ``top'' of the graph in
transformation rules.  

All other vertices in $[P]$ is wrapped in one of two functions $v(\_X)$ or $t(\_X)$. Any vertex equipped with the function $v$
represents a name with the value of $\_X$. On the other hand, vertices equipped
with $t$ describes the behaviour of the corresponding process or represents an
implementation detail that will prove useful later. More
precisely, the argument of $t$ can represent input/output operators
$t(in)/t(out)$, or parallel/sum composition operators $t(p)/t(s)$.

\looseness=-1
We allow vertices with the terms $t(p)$ or $t(s)$ to have a degree less
than $3$ as seen in \figname \ref{fig:syntax-tree-nfp}. In this case, the
vertices do not actually represent a corresponding parallel or sum
composition, but are useful implementation details when implementing the
reduction mechanism with graph transformation. Any input or output
operator is followed by a vertex
with the term $t(p)$, regardless of $t(p)$ representing a parallel composition or not.
Moreover, every vertex with the term $t(p)$, is either a leaf or immediately
followed by a vertex with the label $t(s)$. 

Any prefix operator points to a name corresponding to the
received or sent channel. Since the graphs considered here are not embedded, and
hence do not have an order on the neighbourhood of a vertex, we label an edge
emanating from a prefix operator with ``sync'' or ``arg'' whether the
corresponding name is used for synchronization or as the argument for the
operator.
If a name is both the channel and argument of an operator, then the corresponding edge is equipped with ``arg-sync'' since \MOD does not support parallel edges.

The vertices corresponding to names of
the process are shared among vertices corresponding to input and output
operators. In this way, if an input and output operator synchronize on the same
name, it is illustrated in the resulting graph by both of them pointing to the
vertex corresponding to that name. For example, the outermost input
and output operator of
the process encoded in \figname \ref{fig:syntax-tree-nfp} both synchronize on
the free name $x$, and hence they both contain an edge to the vertex with the term $v(x)$.

Many of the vertices found in $[P]$ are implementation details, that are useful
for simulating process reductions, but makes it difficult to interpret the
behaviour of $P$ simply by looking at $[P]$. In this regard, by default,
\modCalc filters away such implementation details, and instead depicts a
simplified version of the encoding illustrated in \figname
\ref{fig:syntax-simplified}.

\section{Computing Execution Spaces}
\begin{figure}[tbp]
\centering	
		\includegraphics[width = 1\linewidth]{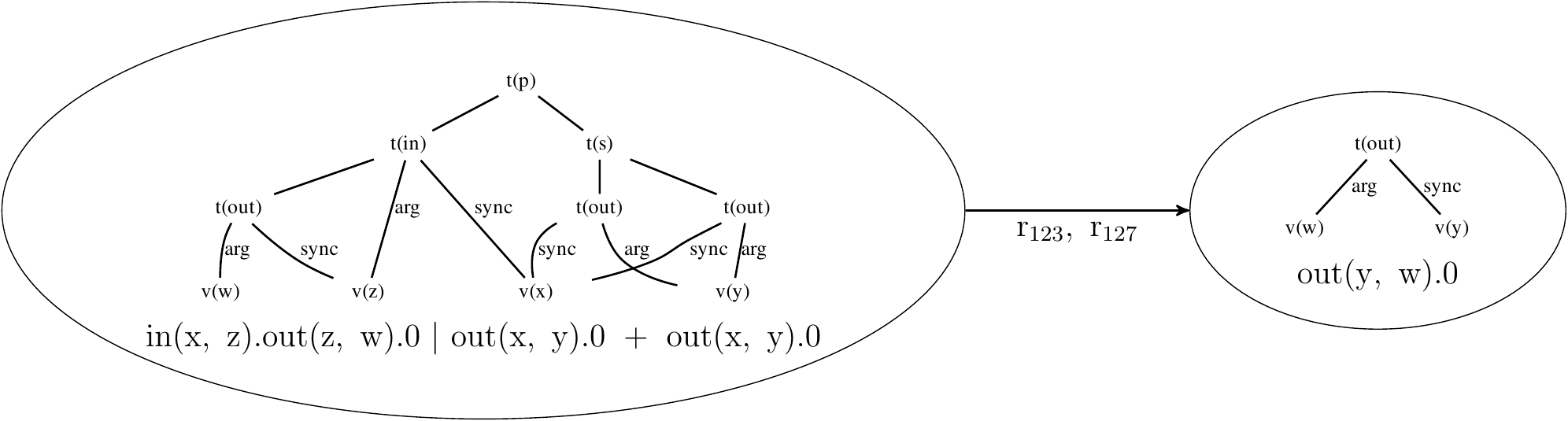}
		\caption{The execution space for $P = \piIn{x}{z}.\piOut{z}{w}\piPar
	(\piOut{x}{y} + \piOut{x}{y})$.}
	\label{fig:reduction-dg-clean}
\end{figure}

\begin{figure}[tbp]
\centering	
		\includegraphics[width = 1\linewidth]{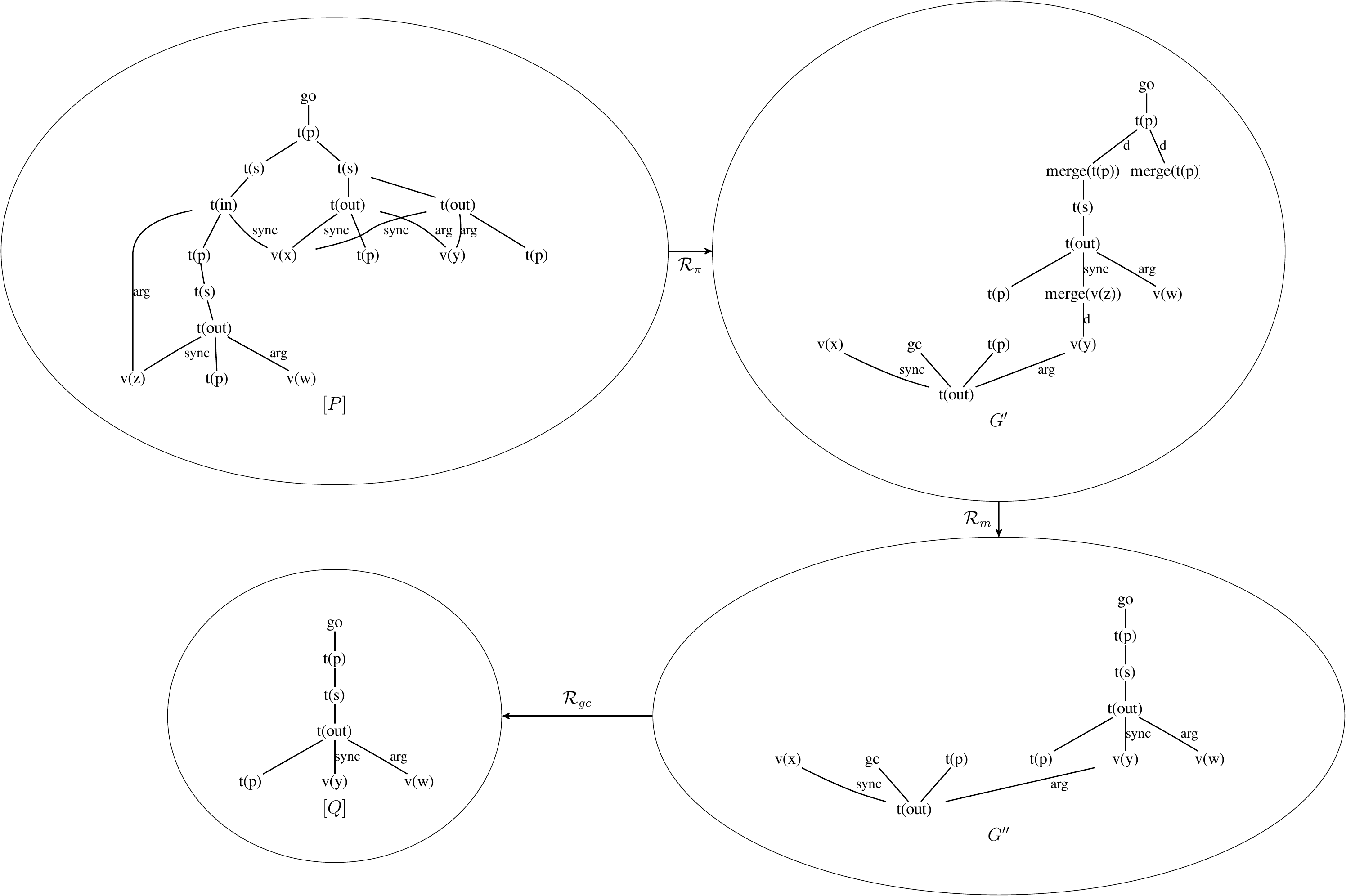}
		\caption{The application of each strategy, $\sysPi$,
		$\sysM$, $\sysGc$, to simulate the reduction $P \rightarrow Q$.}
	\label{fig:reduction-transitions}
\end{figure}

Suppose we were given the process $P = \piIn{x}{z}.\piOut{z}{w}\piPar
(\piOut{x}{y} + \piOut{x}{y})$ and we are able to encode it into its graph
equivalent $[P]$. Supplied with $[P]$, we want to compute the execution space of
$P$.
In \modCalc we can write the following:
\\
\begin{lstlisting}
x, y, z, w = names("x y z w")
p1 = Process().input(x, z).output(z,w)
p2 = Process().output(x, y) + Process().output(x, y)
P = (p1 | p2)
exec_space = ReductionDG(P)
exec_space.calc()
exec_space.print()
\end{lstlisting}
Line 1--4 specifies the process $P$ as explained in the previous section. Line
5--6 computes all possible reductions, while the result is printed in line 7.
The result is illustrated in \figname \ref{fig:reduction-dg-clean}.
The execution space of $P$ is illustrated as a directed graph $G$ where each vertex in $G$ corresponds to a process encoding,
$[Q]$, derived from reductions on $P$, while each edge, $([Q], [R])$, represents the reduction $Q \rightarrow R$, for some processes $Q$ and $R$.
For instance, \figname \ref{fig:reduction-transitions} contains the transformation that encodes the reduction $\piIn{x}{z}.\piOut{z}{w}\piPar
(\piOut{x}{y} + \piOut{x}{y}) \to \piOut{y}{w}$.

In practice, the reduction step is simulated as three distinct strategies.
The first strategy, $\sysPi$, takes care of the actual reduction ($\to$),
while the two other strategies, $\sysGc$, $\sysM$, functions as ``house-cleaning'',
ensuring that the transformed graphs represent process encodings. 

\begin{figure}[tbp]
\summaryRuleSpan{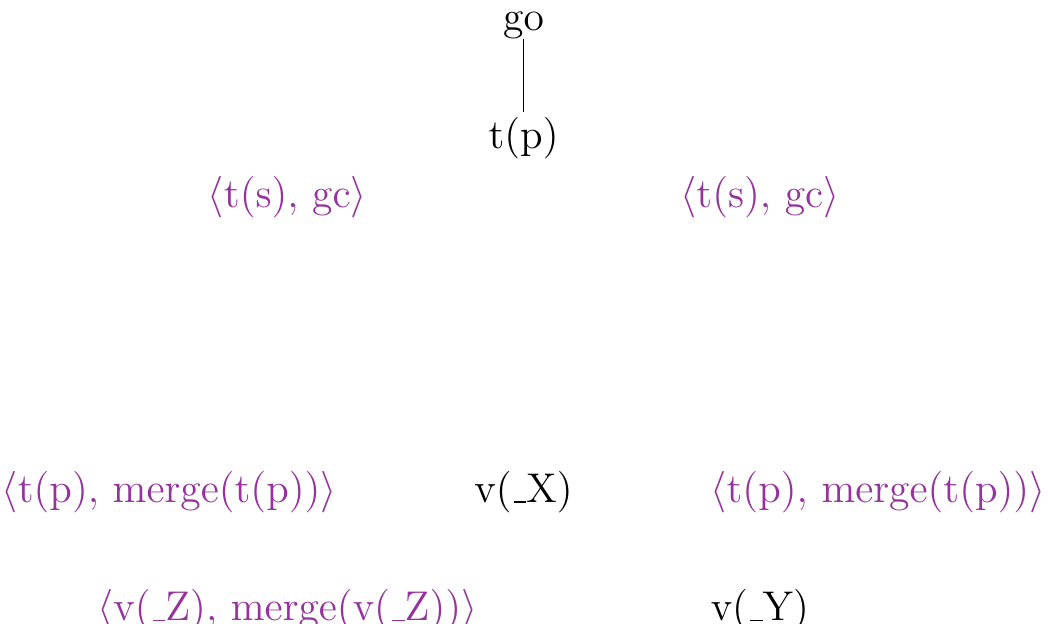}
	\caption{The rule for synchronization in $\sysPi$ when channel
	and argument differ, represented as the span $L \leftarrow K \rightarrow
	R$. Terms starting with an underscore represents variables, e.g., the match
	in $L$ is searching for a name ``v(\_X)'', where the actual name is
	unspecified and contained in ``\_X''. The variable can then be used in $R$.
	Matched vertices and edges coloured
	blue in $L$ will be modified by the
	rule while vertices and edges coloured black will be in the context $K$
	of the rule.
	We see, e.g., that the vertex with the term ``go'' remains unchanged,
	while the vertex with the term ``t(in)'' in $L$ is deleted in $R$.
	Vertices whose terms will be relabelled will be coloured purple in
	the context.
	We see that vertices with the term ``t(s)'' in $L$ will be
	renamed to ``gc'' in $R$, represented in the context with ``$\mathrm{\langle t(s), gc\rangle}$''.
	Finally, vertices marked for coalescing will be wrapped in a term called ``merge'' and connected to the
	vertex they should be coalesced into. This is seen by the renaming of the name
	``v(\_Z)'' to ``merge(v(\_Z))'' and connecting it to ``v(\_Y)'', representing the
	name ``\_Y'' being bound to ``\_Z'' in the simulated reduction.
	}
	\label{fig:reduction-rule}
\end{figure}

The strategy $\sysPi$ consists of two rules,
one of which is depicted in \figname \ref{fig:reduction-rule}.
The rule in \figname \ref{fig:reduction-rule} simulates the reduction $(\piIn{x}{y}.P
+ M) \piPar (\piOut{x}{w}.Q + R) \rightarrow P\{y/w\} \piPar Q$. The 
match of the rule searches for the parallel composition on the outermost prefix
operators, i.e., a vertex labelled $t(p)$ attached to the root. Recall that every
vertex corresponding to a parallel composition is followed by the sum
composition vertex with the label $t(s)$. One of these branches of the
sum composition vertex is matched such that we find an input and output operator
vertex both synchronizing on the same name.

When a match is found, the reduction is then simulated by deleting the prefix
operators and marking their continuations (which are always followed by a
parallel composition vertex) to be coalesced into the ``top'' parallel composition
operator. Additionally, the name $y$ is now bound to $w$, which is simulated by
merging the corresponding vertices. 

An example of applying $\sysPi$ to an encoded process is shown in \figname
\ref{fig:reduction-transitions} as the direct derivation $[P] \Rightarrow G'$,
where $[P]$ describes the process defined in the start of this section.
Note, that the rules are constructed such that when applying a
rule any resulting connected component not containing the root vertex will
never be part of the corresponding process encoding and hence such components
are not depicted
here. The input operator in $P$ can synchronize on the
output operator in both branches of $P$, and similarly there exists
two matches of the rule depicted in \figname \ref{fig:reduction-rule} when applied to
$[P]$. Either match of the rule, however, 
leads to a graph isomorphic to $G'$. The graph $G'$  does not
represent an actual encoding of a process. The reason for this is two-fold:
First, we have simply marked vertices that must be coalesced, but not actually
coalesced them yet, due to the DPO approach in \MOD only supporting injective morphisms, and second, the encoding of any branch not chosen, in this
case one of the output operators of $P$, must be deleted. The coalescing of
vertices is done by $\sysM$, obtaining the graph $G''$ from $G'$ in \figname
\ref{fig:reduction-transitions}, while the deletion of non-chosen summation
branches is done by $\sysGc$, finally resulting in the graph $[Q]$ obtained
from $G''$ in \figname \ref{fig:reduction-transitions}. For more details on
$\sysM$ and $\sysGc$ see Appendix \ref{app:coalescion-gc}.

It was shown in \cite{gadducci2007graph} that for the resulting graph $[Q]$
there exists some process $Q$ with the encoding $[Q]$. Let $[P]
\xRightarrow{\sysPi}_* [Q]$ be the sequence of derivations obtained by applying
$\sysPi$, $\sysM$, and $\sysGc$ on $[P]$ as illustrated in \figname
\ref{fig:reduction-transitions} transforming the encoded process $[P]$ into
some encoded process $[Q]$.
Then, it was additionally shown that given two processes $P$ and $Q$, the
reduction $P \rightarrow Q$ exists iff $[P] \xRightarrow{\sysPi}_* [Q]$ exists.

The actual derivation graph obtained from computing every possible transition $[P]
\xRightarrow{\sysPi}_* [Q]$, involves many direct derivations, as is evident
from \figname \ref{fig:reduction-transitions}, that does not correspond to any
specific state of a process, but are necessary implementation details to
transition from one process state to another. 

Equipped with the
derivations illustrated in \figname \ref{fig:reduction-transitions}, however, we can
identify all direct derivations between actual graphs corresponding to
processes. We can then use such direct derivations to create a new abbreviated derivation
graph, representing our execution space, where any sequence of derivations $[P]
\xRightarrow{\sysPi}_* [Q]$ is modelled as an edge.
The result is an execution space as
illustrated in \figname \ref{fig:reduction-dg-clean}, as first introduced
in the start of this section.

\section{A Final Example}
Suppose we want to model the following behaviour:
\begin{itemize}[nosep,noitemsep]
	\item a patient stumbles into a hospital with two doctors; Dr.~Jekyll and Mr.~Hyde;
	\item the patient will be treated by the first doctor the patient comes across;
	\item Dr.~Jekyll will cure the patient while Mr.~Hyde will kill the patient.
\end{itemize}

The behaviour can be simulated with the expression $Hospital = P \piPar J \piPar H$, for
the process $\phi$ defined by:
\begin{align*}
	P  &=_\phi \piOut{stumble}{name}.\piIn{name}{d}.P' &
	P' &=_\phi \piIn{kill}{x} + \piIn{cure}{x}.P\\
	J &=_\phi \piIn{stumble}{n}.\piOut{n}{jekyll}.\piOut{cure}{j}&
	H &=_\phi \piIn{stumble}{n}.\piOut{n}{hyde}.\piOut{kill}{h}
\end{align*}

We refer to Appendix \ref{app:recursive-process}
for details about the encoding of recursive processes and how to compute the execution spaces including
recursive processes.

 The arguments are not given for the recursive processes,
as they are just the
set of free names of the respective process. A patient walks into the hospital
and gives their name to either Jekyll or Hyde via the stumble channel. The doctor
then sends his own name back to the patient as a greeting via the patients name as
a channel (note, this serves no practical purpose for the model). Finally, the
patient is either cured or killed,
depending on the doctor received. If cured, $P$ is called recursively
simulating that the patient can get sick and stumble into the hospital again. If the patient
died, we reached a deadlock as there is no cure for death. 
Note, many of the arguments sent over the channels are not used, but are simply
there for illustrative purposes.

We can model the above example in \modCalc as follows:
\begin{lstlisting}
s, n, pn, j, h, cu, ki, d, x = names("s n pn j h cu ki d x")
free_names = [s, n, ki, cu] 

rp = RecursiveProcess()

P = Process().output(s, n).input(n, d).call("Pp", free_names)
rp.add("P", free_names, P)

Pp = Process().input(ki, x) + Process().input(cu, x).call("P", free_names)
rp.add("Pp", free_names, Pp)

J = Process().input(s, pn).output(pn, j).output(cu, j).call("J", free_names)
rp.add("J", free_names, J)

H = Process().input(s, pn).output(pn, h).output(ki, h).call("H", free_names)
rp.add("H", free_names, H)

Hospital = P | J | H

exec_space = ReductionDG(Hospital, rp)
exec_space.calc()
exec_space.print()
\end{lstlisting}

\begin{figure}[tbp]
	\centering
	\includegraphics[width=1\textwidth]{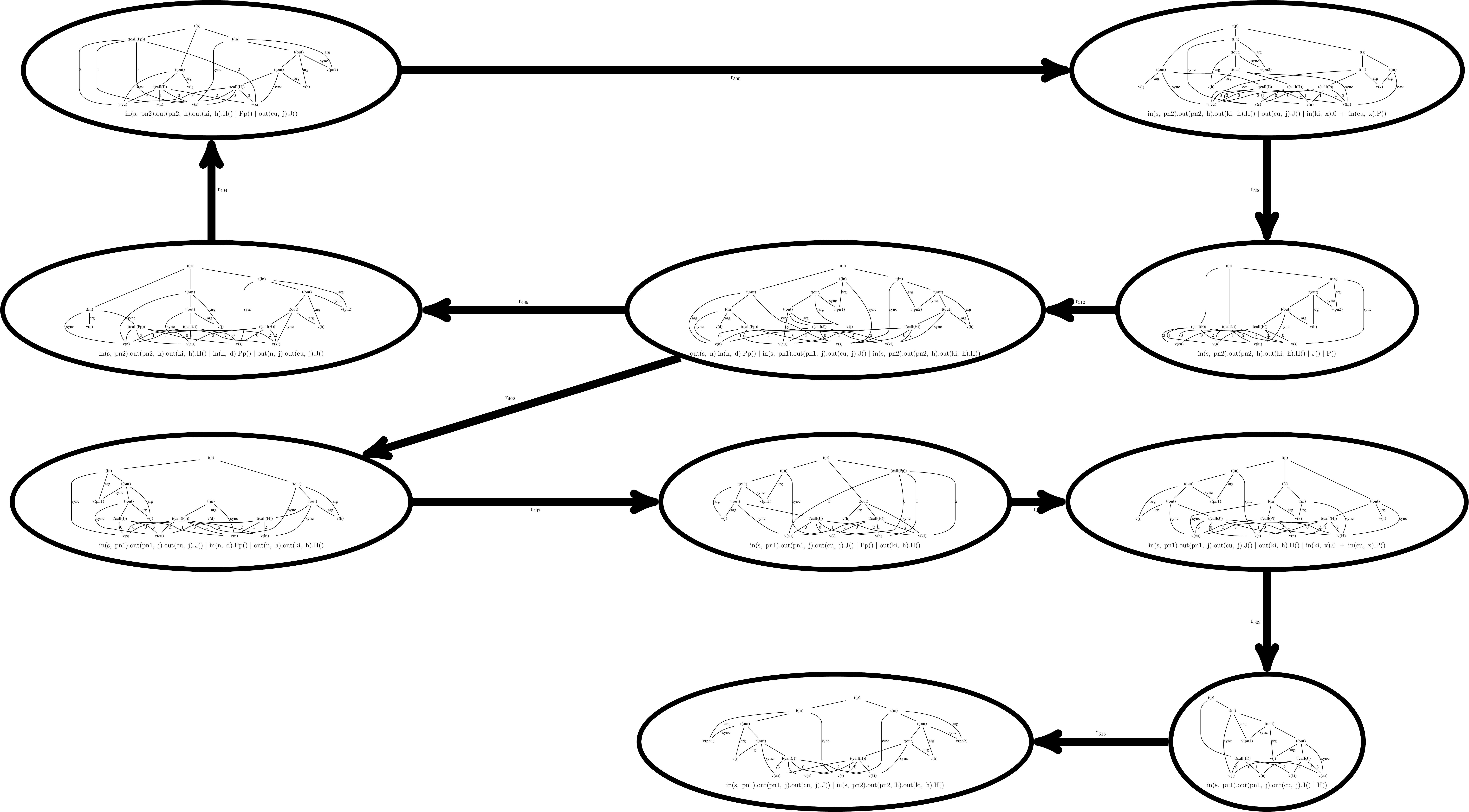}
	\caption{ The execution space for $Hospital$.
	The start of the simulation, i.e., $[Hospital]$, is the single vertex with a degree of 3.
	The vertex for $[Hospital]$ contains two out-edges corresponding to the patient $P$ either
	synchronizing on Dr.~Jekyll $J$ or Mr.~Hyde $H$.
	If $P$ synchronizes with $J$, we see that we end up in a cycle, representing $P$ being cured and
	called recursively. On the other hand synchronizing on $H$, leads to a vertex
	with no out-edges, and hence no reductions are possible from the
	corresponding encoding, representing the patient being killed off and we
	are left with the encoded process $[H\piPar J]$.}
		\label{fig:hospital}
\end{figure}

Note, we have shortened the names to simple single letter names for illustrative purposes.
The resulting execution space, representing the simulation of $Hospital$, is
illustrated in \figname \ref{fig:hospital}.
Although, it might be difficult to parse the graph, the expected behaviour
should be clear from the graph: if the patient is treated by Dr.~Jekyll, $P$ is called recursively and as a result the behaviour is modelled as
the only cycle in the derivation graph. If on the other hand the patient was
killed off then $P$ is terminated. Hence the behaviour is modelled in the
graph as a path that end in a vertex with no outgoing edges (reached a
deadlock).

\section{Conclusion}
We have introduced \modCalc, available as a web front-end at \url{http://cheminf.imada.sdu.dk/epim};
a tool using graph transformation for computing
execution spaces of $\pi$-calculus processes. In this regard we presented a brief
overview of the encodings and transformations involved. Practically,
execution spaces can be used for basic analysis of processes such
as determining liveness.  
Since \modCalc is directly embedded into \MOD, it would be possible
to use the range of features provided by the \MOD framework, notably \cite{graphCanon}.
Since structural congruence can be determined directly from graph isomorphism of the
encoded processes, canonicalization of processes is given for free.
\MOD provides a framework for stochastic simulations \cite{gillespie}, suggesting the possibilities of using a version of the encoding of $\pi$-calculus presented
here to simulate stochastic $\pi$-calculus processes~\cite{cardelli2013stochastic}.

The type of DPO supported by \MOD does not allows rewriting rules that perform bulk duplication or deletion of subgraphs. This means that duplicating or deleting processes cannot be implemented with a single rewrite operation. On one hand, this elicits the cost of process duplication or recursion often ignored in process calculi, on the other, it suggests to explore the use of \MOD with ``resource aware calculi'' like linear variations of the $\pi$-calculus \cite{DBLP:journals/toplas/KobayashiPT99,DBLP:conf/concur/CairesP10,DBLP:journals/pacmpl/KokkeMP19}.

\bibliography{biblio}{}

\begin{thebibliography}{10}

\bibitem{strat:14}
Jakob~L. Andersen, Christoph Flamm, Daniel Merkle, and Peter~F. Stadler.
\newblock Generic strategies for chemical space exploration.
\newblock {\em Int. J. Comp. Bio. and Drug Design}, 2014.

\bibitem{andersen2016software}
Jakob~L. Andersen, Christoph Flamm, Daniel Merkle, and Peter~F. Stadler.
\newblock A software package for chemically inspired graph transformation.
\newblock In {\em Graph Transformation - 9th International Conference, {ICGT}
  2016, Proceedings}, volume 9761 of {\em Lecture Notes in Computer Science},
  pages 73--88. Springer, 2016.

\bibitem{Andersen17}
Jakob~L. Andersen, Christoph Flamm, Daniel Merkle, and Peter~F. Stadler.
\newblock Chemical transformation motifs --- modelling pathways as integer
  hyperflows.
\newblock {\em IEEE/ACM Transactions on Computational Biology and
  Bioinformatics}, 2018.
\newblock in press, early access online.

\bibitem{gillespie}
Jakob~L. Andersen, Christoph Flamm, Daniel Merkle, and Peter~F. Stadler.
\newblock Rule-based gillespie simulation of chemical systems.
\newblock In {\em 3rd Workshop on Verification of Engineered Molecular Devices
  and Programs, {VEMDP} 2016, Proceedings}. Springer, 2018.

\bibitem{graphCanon}
Jakob~L. Andersen and Daniel Merkle.
\newblock A generic framework for engineering graph canonization algorithms.
\newblock In {\em 2018 Proceedings of the 20th Workshop on Algorithm
  Engineering and Experiments (ALENEX)}, pages 139--153, 2018.

\bibitem{bog2006tool}
Anja Bog and Frank Puhlmann.
\newblock A tool for the simulation of $\pi$-calculus systems.
\newblock {\em Open. BPM}, 2006.

\bibitem{DBLP:conf/concur/CairesP10}
Lu{\'{\i}}s Caires and Frank Pfenning.
\newblock Session types as intuitionistic linear propositions.
\newblock In {\em {CONCUR}}, volume 6269 of {\em Lecture Notes in Computer
  Science}, pages 222--236. Springer, 2010.

\bibitem{cardelli2013stochastic}
Luca Cardelli and Radu Mardare.
\newblock Stochastic pi-calculus revisited.
\newblock In {\em International Colloquium on Theoretical Aspects of
  Computing}, pages 1--21. Springer, 2013.

\bibitem{kappa}
Vincent Danos, J{\'{e}}r{\^{o}}me Feret, Walter Fontana, Russell Harmer, and
  Jean Krivine.
\newblock Rule-based modelling of cellular signalling.
\newblock In {\em {CONCUR} 2007 - Concurrency Theory, 18th International
  Conference, Proceedings}, volume 4703 of {\em Lecture Notes in Computer
  Science}, pages 17--41. Springer, 2007.

\bibitem{ehrig1973graph}
Hartmut Ehrig, Michael Pfender, and Hans~J{\"u}rgen Schneider.
\newblock Graph-grammars: An algebraic approach.
\newblock In {\em 14th Annual Symposium on Switching and Automata Theory
  ({SWAT} 1973)}, pages 167--180. IEEE, 1973.

\bibitem{gadducci2007graph}
Fabio Gadducci.
\newblock Graph rewriting for the $\pi$-calculus.
\newblock {\em Mathematical Structures in Computer Science}, 17(3):407--437,
  2007.

\bibitem{habel2001double}
Annegret Habel, J{\"u}rgen M{\"u}ller, and Detlef Plump.
\newblock Double-pushout graph transformation revisited.
\newblock {\em Mathematical Structures in Computer Science}, 11(5):637--688,
  2001.

\bibitem{bionetgen}
Leonard~A. Harris, Justin~S. Hogg, Jos{\'{e}}~Juan Tapia, John A.~P. Sekar,
  Sanjana Gupta, Ilya Korsunsky, Arshi Arora, Dipak Barua, Robert~P. Sheehan,
  and James~R. Faeder.
\newblock Bionetgen 2.2: advances in rule-based modeling.
\newblock {\em Bioinformatics}, 32(21):3366--3368, 2016.

\bibitem{DBLP:journals/toplas/KobayashiPT99}
Naoki Kobayashi, Benjamin~C. Pierce, and David~N. Turner.
\newblock Linearity and the pi-calculus.
\newblock {\em {ACM} Trans. Program. Lang. Syst.}, 21(5):914--947, 1999.

\bibitem{DBLP:journals/pacmpl/KokkeMP19}
Wen Kokke, Fabrizio Montesi, and Marco Peressotti.
\newblock Better late than never: a fully-abstract semantics for classical
  processes.
\newblock {\em {PACMPL}}, 3({POPL}):24:1--24:29, 2019.

\bibitem{konig2000graph}
Barbara K{\"o}nig.
\newblock A graph rewriting semantics for the polyadic pi-calculus.
\newblock In {\em Proc. of GT-VMT’00 (Workshop on Graph Transformation and
  Visual Modeling Techniques)}, pages 451--458, 2000.

\bibitem{milner1999pibook}
Robin Milner.
\newblock {\em Communicating and mobile systems - the Pi-calculus}.
\newblock Cambridge University Press, 1999.

\bibitem{sangiorgi2001pibook}
Davide Sangiorgi and David Walker.
\newblock {\em The Pi-Calculus - a theory of mobile processes}.
\newblock Cambridge University Press, 2001.

\bibitem{Victor102920}
Bj{\"o}rn Victor and Faron Moller.
\newblock The mobility workbench : A tool for the pi-calculus.
\newblock In {\em Proceedings of CAV'94 :}, pages 428--440, 1994.

\end{thebibliography}
\bibliographystyle{plain}

\newpage

\appendix

\section{Graph Morphisms with First-Order Terms}\label{app:graph-morphisms}
\subsection{First-Order Terms and Unification\protect \footnote{meant for
    inclusion in conference proceedings}}
\label{sec:terms}
\label{sec:term:moreGeneral}
In \MOD vertices and edges of graphs are labelled first-order terms in the
style known from Prolog, and syntactic unificiation is used during label matching.

Let $\mathcal{F}$ denote a set of function symbols and $\mathcal{V}$ a set of variable symbols.
The set of first-order terms $\mathcal{T}(\mathcal{F}, \mathcal{V})$ is then defined as the smallest set such that:
\begin{itemize}
\item all variables are terms, i.e., $V\subseteq \mathcal{T}(\mathcal{F}, \mathcal{V})$,
\item and for all arities $n\in \mathbb{N}_0$, function symbols $f\in \mathcal{F}$, and terms $t_1, \dots, t_n\in \mathcal{T}(\mathcal{F}, \mathcal{V})$,
    we have $f(t_1, \dots, t_n)\in \mathcal{T}(\mathcal{F}, \mathcal{V})$
\end{itemize}

A substitution $\sigma$ is a mapping $\{v_1\mapsto t_1, \dots, v_{|\sigma|}\mapsto t_{|\sigma|}\}$ of variables to terms.
It can be applied to a term $t$ by replacing all occurrences of each $v_i$ in $t$ with the corresponding term $t_i$.
In short we write the application of a substitution as $\sigma(t)$.
If a substitution only maps variables into variables such that it is a bijection, then it is called a \emph{renaming}.
In algorithmic contexts we also refer to a substitution as a set of \emph{variable bindings}.

Between two terms $t_1$ and $t_2$ we use the following relations:
\begin{itemize}
\item Equality of $t_1$ and $t_2$ is written $t_1 = t_2$.
For example $\str{f}(X) = \str{f}(X)$, but $\str{f}(X) \neq\str{f}(Y)$ when $X$ and $Y$ are different variable symbols.
\item $t_2$ is \emph{at least as general} as $t_1$, written as $t_1\moreSpecial t_2$ or $t_2\moreGeneral t_1$, if there exist a substitution $\sigma$ such that $\sigma(t_2) = t_1$.
For example, $\texttt{f}(A, B)\moreGeneral \texttt{f}(X, X)$ because the terms become equal when $A$ and $B$ are replaced with $X$ in the left term.
\item $t_1$ and $t_2$ are \emph{isomorphic}, written $t_1\iso t_2$, if both $t_1\moreSpecial t_2$ and $t_2\moreSpecial t_1$.
Equivalently, $t_1\iso t_2$ if there exists a renaming $\sigma$ such that $t_1 = \sigma(t_2)$.
Thus $\texttt{f}(A, B)\iso \texttt{f}(X, Y)$ and $\texttt{f}(A, B)\iso \texttt{f}(B, A)$,
but $\texttt{f}(A, B)\not\iso\texttt{f}(Z, Z)$.
\item $t_1$ and $t_2$ are \emph{unifiable}, written $t_1\unifiable t_2$, if there exist a substitution $\sigma$ such that $\sigma(t_1) =\sigma(t_2)$.
Such a substitution is called a \emph{unifier} of $t_1$ and $t_2$.
For example, a unifier for $\str{f}(X, \str{g}(Y))$ and $\str{f}(Z, Z)$ is $\{X\mapsto \str{g}(Y), Z\mapsto \str{g}(Y)\}$.
\end{itemize}

The \emph{most general unifier} (mgu) of two terms $t_1$ and $t_2$ is the unifier $\sigma$ such that for any other unifier $\sigma'$, we have $\sigma(t_1) \moreGeneral \sigma'(t_1)$.
That is, the unifier $\sigma$ produces the most general terms of all unifiers.
For example, $\sigma' = \{X\mapsto \str{g}(\str{a}), Y\mapsto \str{a}, Z\mapsto \str{g}(\str{a})\}$ is not the mgu of $t_1 = \str{f}(X, \str{g}(Y))$ and $t_2 = \str{f}(Z, Z)$ because there is another unifier $\sigma = \{X\mapsto \str{g}(Y), Z\mapsto \str{g}(Y)\}$, and $\sigma(t_1) = \str{f}(\str{g}(Y), \str{g}(Y))$ which is more general than $\sigma'(t_1) =  \str{f}(\str{g}(\str{a}), \str{g}(\str{a}))$.

Deciding if the three relations $t_1\iso t_2$, $t_1\moreSpecial t_2$, and $t_1\unifiable t_2$ hold can be seen as different levels of pattern matching with isomorphism for exact matching, specialisation/generalisation for one-sided matching, and unification for two-sided matching.
Assuming we have an algorithm for computing the most general unifier $\sigma$ of $t_1$ and $t_2$, if it exists,
we can from $\sigma$ also see if the terms are isomorphic by checking if $\sigma$ is a renaming.
A variant of such a unification algorithm can also decide if $t_1\moreGeneral t_2$ by performing the unification but disallowing binding of variables in $t_2$.

\subsection{Labelled Graph Morphisms\protect\footnote{not meant for inclusion in
    conference proceedings}}
A labelled graph is a tuple $G = (V_G, E_G, l^V_G, l^E_G)$, where $(V_G, E_G)$ is the underlying graph, $l^V_G\colon V_G\rightarrow \Omega_V$ is the function labelling vertices with elements from some set $\Omega_V$, and
$l^E_G\colon E_G\rightarrow \Omega_E$ is the function labelling edges with elements from some set $\Omega_E$.
A graph morphism $m\colon G\rightarrow H$ on labelled graphs induces the label associations 
\begin{align*}
A_V(m) &= \{\langle l_G^V(v), l_H^V(m(v))\rangle \mid v\in V_G\}        \\
A_E(m) &= \{\langle l_G^E(e), l_H^V(m(e))\rangle \mid e\in E_G\}
\end{align*}
Depending on the structure of $\Omega_V$ and $\Omega_E$ we can then define different kinds of morphisms.
For example, if the labels are character strings we simply require that all the associated labels from the morphism are equal, i.e., $s_1 = s_2, \forall (s_1, s_2)\in A_V(m)\cup A_E(m)$.

As a generalisation we label vertices and edges with first-order terms, and direct equality of the terms is not always desired.
Given two graphs $G$ and $H$ labelled with first-order terms, and a morphism $m\colon G\rightarrow H$.
Let $t_G$ and $t_H$ be the aggregate terms
\begin{align*}
    t_G = \str{assoc}(&
            l_V(v_1), l_V(v_2), \dots, l_V(v_{|V_G|}),  \\  
        &   l_E(e_1), l_E(e_2), \dots, l_E(e_{|E_G|})
        )   \\  
    t_H = \str{assoc}(&
            l_V(m(v_1)), l_V(m(v_2)), \dots, l_V(m(v_{|V_G|})), \\
        &   l_E(m(e_1)), l_E(m(e_2)), \dots, l_E(m(e_{|E_G|}))
        )
\end{align*}
for some arbitrary ordering of $V_G$ and $E_G$, and a new function symbol \str{assoc}.
If $m$ is an isomorphism and $t_G\iso t_H$, then $G$ and $H$ not only have the same graph structure, but the labelling is the same, except for renaming of variables.
However, if $m$ is a monomorphism we can define different levels of pattern matching by checking if either $t_G\iso t_H$, $t_G\moreGeneral t_H$, or $t_G\unifiable t_H$.
Requiring $t_G\iso t_H$ can be interpreted as a check for exact substructure, while $t_G\moreGeneral t_H$ can be used to check if a pattern is less restrictive than another.
As noted in above, deciding the relation $\moreGeneral$ can be done via one-sided unification.

\section{Coalescion and Garbage Collection Rules\protect\footnote{not meant
    for inclusion in conference proceedings}}\label{app:coalescion-gc}
For coalescing of vertices we employ the strategy $\sysM$.
It was shown in \cite{habel2001double} how to coalesce vertices to simulate a DPO approach
with arbitrary morphisms from the left-hand to the right-hand side of a rule in
a DPO approach that only allows for injective morphisms between them. Here, we
alter the approach slightly. A vertex is mapped for coalescing by wrapping the vertex in question 
with the term ``merge'', such as the vertex ``merge(t(p))'' found in $G'$.
Additionally, any vertex marked for coalescing is connected to the vertex it
should be coalesced into by an edge labelled with the term ``d'', e.g., the
parallel composition operator vertex with the term ``t(p)'' is connected to the vertices with
the term ``merge(t(p))'' in $G'$. The strategy $\sysM$, simply
moves the neighbourhood of any vertex marked for coalescing into the vertex
it should be coalesced with, and then finally deletes the vertex marked for
coalescing when its neighbourhood is empty. Applying $\sysM$ to $G'$ results in
the graph $G''$ shown in \figname \ref{fig:reduction-transitions}. Obtaining $G''$ from
$\sysM$, involves a series of direct derivations, however, $\sysM$ is strongly
confluent and hence $\sysM$ will always derive the graph $G''$ from $G'$ by
continuously applying $\sysM$ on the result until termination.

\begin{figure}[tbp]
	\centering
	\begin{minipage}{0.8\textwidth}
\summaryRuleSpan{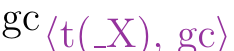}
\summaryRuleSpan{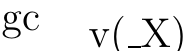}
		
	\end{minipage}
	\caption{
		The reduction cleanup rules of $\sysGc$.
		}
		\label{fig:reduction-rule-cleanup}
\end{figure}

To remove the encoding of the corresponding branch of a summation operator, we
first mark the vertex representing the summation operator to be removed with
the term ``gc'', such that we can garbage collect it later. See $G''$ for an example.
To ``garbage-collect'' summation branches, 
we introduce the strategy
$\sysGc$ containing the two rules illustrated in \figname \ref{fig:reduction-rule-cleanup}.

The rules simply propagates the ``gc'' term to all vertices equipped with the
function $t$ adjacent to a vertex marked for garbage collection while removing
all edges adjacent to a garbage collected vertex. Again, $\sysGc$ is strongly
confluent, and hence the sequence of derivations leads to the unique graph, e.g.,
$[Q]$ from $G''$, where every branch of the sum composition
operator is deleted in the corresponding graph.

\section{Recursive Processes\protect\footnote{not meant for inclusion in
    conference proceedings}}\label{app:recursive-process}
In this section we show how to extend our framework to include recursive
processes, i.e., processes defined using constant invocation.

For the inclusion of recursive processes we extend the notion of a process
expression as defined in \secname \ref{sec:pi-calc} with the term $A(x_1, \dots,
x_n)$, where $x_1,\dots, x_n$ is a sequence of names and we let $fn(A(x_1, \dots,
x_n)) = \{ x_1, \dots, x_n \}$.

A recursive process $\phi$ is then a finite set of equations (at most
one for each process identifier $A$) of the following kind:
$$
A(x_1, \dots, x_n) =_{\phi} P_A
$$
for $x_i$ distinct names and $P_A$ process expressions.

Intuitively, a recursive process corresponds to a procedure definition, where
each identifier in a process expression represents a procedure call.

Finally, we extend the reduction semantics of recursive $\pi$-calculus with the
following axiom:
$$
	\inferrule{A(x_1, \dots, x_n) =_\phi P_A}{A(y_1, \dots, y_n) \rightarrow_\phi P_A\{y_1/x_1, \dots, y_n/x_n\}}
$$

\subsection{Encoding Recursive Processes as Graphs}\label{sec:enc-recursive}
Intuitively, a recursive process corresponds to a procedure definition, where
each identifier in a process expression represents a procedure call.

\begin{figure}[tbp]
	\centering
	\subfloat[]{
		\includegraphics[width=0.35\textwidth]{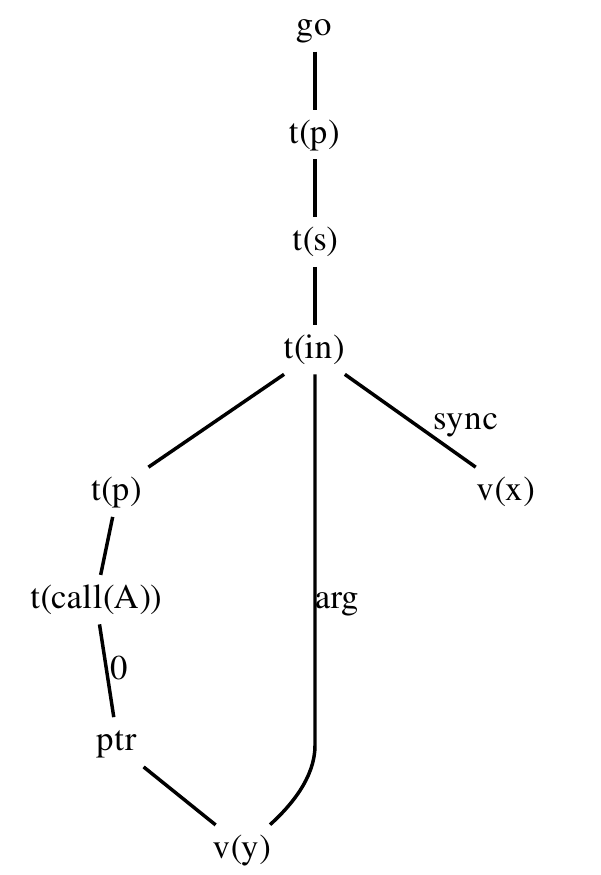}
	\label{fig:encode-A}
		}
	\subfloat[]{
		\includegraphics[width=0.35\textwidth]{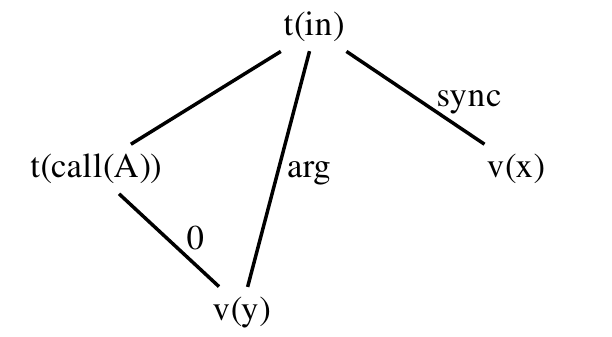}
		\label{fig:encode-A-simplified}
		}
	\caption[]{
		\subref{fig:encode-A} The encoded process $[A(x)]$ for $A(x) =_{\phi} \piIn{x}{y}.A(y)$.
		\subref{fig:encode-A-simplified} The simplified version of $[A(x)]$.
	}
\end{figure}

Defining a recursive process in \modCalc is done as follows:
\begin{lstlisting}
x, y, z = names("x y z")
rp = RecursiveProcess()

A = Process().input(x, y).call("A", [y])
rp.add("A", [x], A)

B = Process().output(x, x).call("B", [x])
rp.add("B", [x], B)
\end{lstlisting}
The above example defines the recursive process $\phi$ defined by:
\begin{align*}
	A(x) &=_\phi x(y).A(y) \\
	B(x) &=_\phi \bar{x}x.B(x)
\end{align*}
We can of course also encode such processes as graphs. The encoded process
$[A(x)]$ is illustrated in \figname \ref{fig:encode-A}. Here, the recursive
invocation of $A(y)$ is modelled by the vertex labelled t(call(A)). The arguments
of the invocation are modelled as edges attached to vertices representing
``pointers'', which in turn
points to the names given as arguments. Since edges are not ordered, we label
each argument edge with their corresponding argument position in the
invocation, i.e., $y$ is given as the first (zero'th) argument to the invocation
of $A(y)$ in \figname \ref{fig:encode-A}. Of course, like before, concepts such as
pointers are implementation details that helps modelling process invocations as
graph transformations, but obscures the actual behaviour of the process. Hence,
by default such implementation details a filtered away when depicting processes
as graphs, and instead $[A(x)]$ is depicted as in \figname
\ref{fig:encode-A-simplified}.

\subsection{Unfolding Recursive Process Calls}
Now, suppose we are given the process $P = A(x) \piPar B(x)$ and the recursive
process $\phi$ defined by:
\begin{align*}
	A(x) &=_\phi \piIn{x}{y}.A(y) \\
	B(x) &=_\phi \piOut{x}{x}.B(x)
\end{align*} 

\begin{figure}[tbp]
	\centering
	\includegraphics[width=1\textwidth]{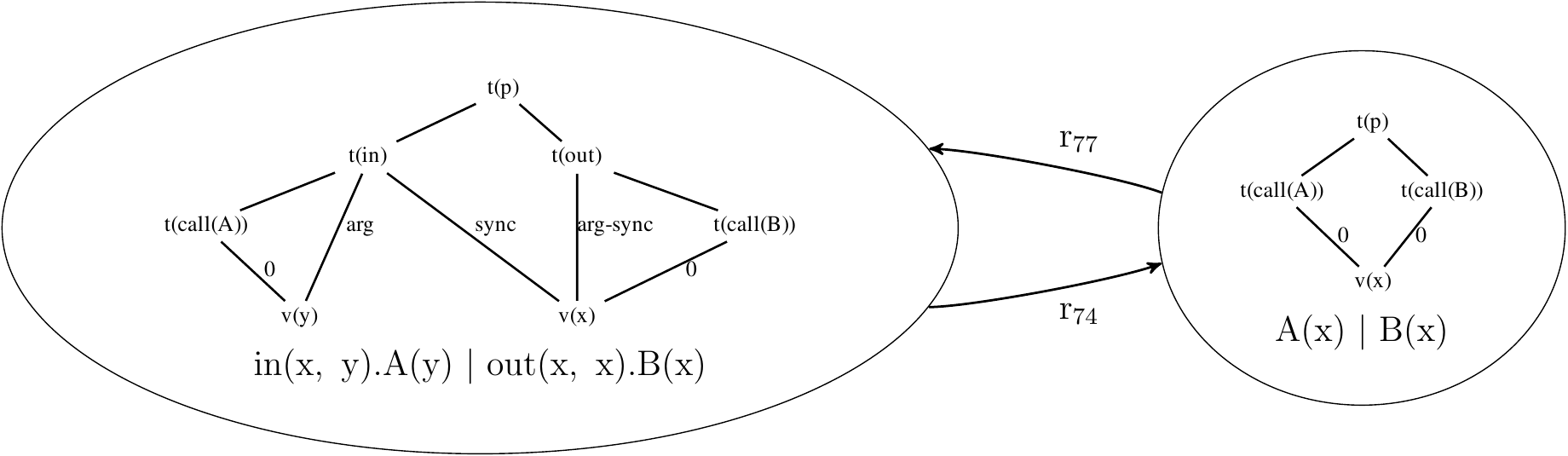}
	\caption{
		The execution space for $P$. First, the call sites are unfolded and
		then synchronized obtaining the original process $P$.
		}
		\label{fig:recursive-reduction-example}
\end{figure}

Computing the execution space for $P$ given a recursive process $\phi$ in \modCalc can be done
as follows:
\begin{lstlisting}
x, y, z = names("x y z")
rp = RecursiveProcess()

A = Process().input(x, y).call("A", [y])
rp.add("A", [x], A)

B = Process().output(x, x).call("B", [x])
rp.add("B", [x], B)

P = A | B
exec_space = ReductionDG(P, rp)
exec_space.calc()
exec_space.print()
\end{lstlisting}
Line 1--10 defines the behaviour of $P$ and $\phi$ while line 11--12 computes the
execution space that is printed in line 13. Note, that the space is computed
exactly as in the previous section, with the exception that the recursive
process $\phi$ is provided in line 11. The resulting execution space is
illustrated in \figname \ref{fig:recursive-reduction-example}.

When including recursive processes, and hence the possibility to include
process invocations, simply simulating the reduction mechanism is not adequate
as illustrated in \figname \ref{fig:recursive-reduction-example}. Before we can
synchronize $A(x)$ and $B(x)$ on $x$ in $P$ we must first expand, or unfold, the
corresponding process calls.

\begin{figure}[tbp]
	\summaryRuleSpan{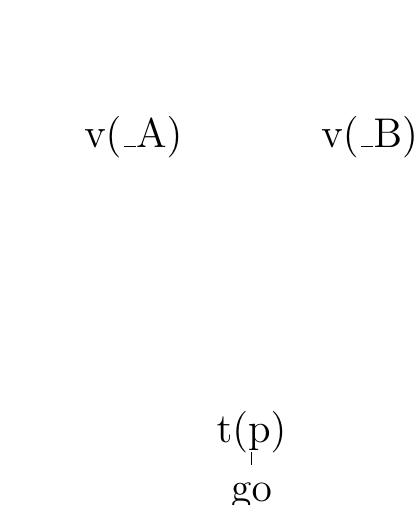}
	\caption{
		The rule for $A(x, y) =_\phi \piOut{x}{y}.A(x, y)$. Note, that the rules
		does not specify actual names but uses the positional arguments stored
		in the edges corresponding to arguments, to correctly expand a process.
		}
	\label{fig:recursive-rule}
\end{figure}

In this regard, we introduce for each recursive process $\phi$ the strategy
$\sysR$ containing an unfolding rule $p^\phi_A$ for each equation $A(x_1, \dots, x_n)
=_\phi P_A$ in $\phi$.  An example is shown
in \figname \ref{fig:recursive-rule} modelling the invocation of $A(x, y)
=_{\phi} \piOut{x}{y}.A(x, y)$. When unfolding $P_A$, we first
match an invocation of a process $P_A$ contained in the outermost parallel
composition of a process, like for $A(x)$ and $B(x)$ in $P = A(x)\ |\ B(x)$.
When a match is found, 
we expand the vertex with the term ``t(call(A))'' such that the expanded graph
correspond to $[P_A]$, but with the free names replaced by the given arguments
at the call site.

Note, \figname \ref{fig:recursive-rule} is a simplification of the
actual rule used for expansion. The depicted rule does not contain
the pointer vertices, as specified in \secname \ref{sec:enc-recursive}, but refers to the argument names directly. Any matching in \MOD
must be injective. As a consequence, if the arguments referred to names
directly, we would need a different rule if two arguments referred to the same
name. Instead, the actual rule matches on the pointers, and then later merges
the pointers into their attached names using $\sysM$, as described in the
previous section. Intuitively, however, the unfolding of process calls can
be thought of as simulating the rule illustrated in \figname \ref{fig:recursive-rule}.

\end{document}